# Online Calibration of Phasor Measurement Unit Using Density-Based Spatial Clustering

Xinan Wang, *Student Member, IEEE*, Di Shi, *Senior Member, IEEE*, Zhiwei Wang, *Member, IEEE*, Chunlei Xu, Qibing Zhang, Xiaohu Zhang, *Member, IEEE*, Zhe Yu, *Member, IEEE*

*Abstract*—Data quality of Phasor Measurement Unit (PMU) is receiving increasing attention as it has been identified as one of the limiting factors that affect many wide-area measurement system (WAMS) based applications. In general, existing PMU calibration methods include offline testing and model based approaches. However, in practice, the effectiveness of both is limited due to the very strong assumptions employed. This paper presents a novel framework for online bias error detection and calibration of PMU measurement using density-based spatial clustering of applications with noise (DBSCAN) based on much relaxed assumptions. With a new problem formulation, the proposed data mining based methodology is applicable across a wide spectrum of practical conditions and one side-product of it is more accurate transmission line parameters for EMS database and protective relay settings. Case studies demonstrate the effectiveness of the proposed approach.

## I. Introduction

PHASOR measurement unit (PMU) is envisioned to be one of the enabling technologies in smart grid, with the promise of massive installation in future power systems. On one hand, most synchrophasor-based applications, especially the mission critical ones, require the measurements to be very reliable and accurate. On the other, although PMU data are expected to be highly accurate, this potential accuracy and reliability are not always achieved in actual field installation due to various causes [1]. It has been observed under many occasions that PMU measurements can have various types of data quality issues. To ensure accurate, reliable and consistent PMU data, there are pressing needs to calibrate PMU to fulfill the claimed performance.

As discussed in [2], the PMU device itself is typically very accurate, but the instrumental channel, where a PMU gets its inputs, is usually much less accurate. In particular, the instrumentation channel (e.g., potential and current transformers) can introduce magnitude and phase angle errors that are magnitudes of orders higher than the typical PMU accuracy. A practically useful calibration method should be capable of handling inaccuracies originated from both sources.

This work is funded by SGCC Science and Technology Program under contract no. 5455HJ160007.
X. Wang, D. Shi, Z. Wang, X. Zhang, and Y. Zhe are with GEIRI North America, 5451 Great America Parkway, Ste. 125, Santa Clara, CA 95054. (Email: di.shi@geirina.net).
C. Xu and Q. Zhang are with State Grid Jiangsu Electric Power Company, 215 Shanghai Road, Gulou, Nanjing, 210024, China.

Previously the Performance and Standards Task Team (PSTT) published a PMU system testing and calibration guide [3]. As discussed and widely accepted in the 2016 NASPI Work Group meeting, PMU data quality efforts need to be implemented to ensure the highest synchrophasor signal quality for applications. The modified IEEE C37.118 standard requires the total vector error (TVE) between a measured phasor and its true value to be well within 1% under steady-state operating conditions [4]. Towards these requirements, many PMU calibration schemes have been proposed. In general, these methods can be divided into two categories based on how they are implemented: offline testing/calibration [5]-[11] and model-based approaches [12]-[16].

The former works by comparing PMU output against standard testing signal(s), using certain types of specialized equipment or systems whose accuracies are at least one level greater than the to-be-tested PMUs. This type of methods requires specialized equipment/system, and due to their offline nature, errors originated from the instrumentation channel cannot be duplicated and compensated.

The latter works by fitting PMU measurements into a mathematical model for fidelity check, assuming parameters of the system/device(s) and the model are known *a priori* and accurate. In [12], the authors present a phasor-data-based state estimator (PSE) that is capable of identifying and correcting bias error(s) in phase angles. This approach assumes the phasor magnitudes and network parameters are both accurate. Paper [13] proposes the idea of a "super calibrator" for substation-level data filtering and state estimation, the input of which includes PMU data, SCADA data, and a detailed 3-phase model of the substation, etc. Despite complexity of the model, the accuracy level of SCADA data adds uncertainty, or even degrades performance of the approach. Paper [1] proposes a calibration-factor-based iterative non-linear solution approach for 3-phase PMU data calibration. Performance of the approach is highly dependent upon accuracy of the 3-phase transmission line parameters in the EMS database. The PMU data calibration approach in [14] again assumes the TL impedances are known to be exact. Papers [15] and [16] attempt to accomplish line parameter estimation and PMU calibration simultaneously, with the assumption that one of the two PMUs generates perfect measurements, which, in practice, is really difficult to tell. The strong assumptions used in existing model-based methods undermine their practicability.

This paper presents a novel data mining based synchrophasor measurement calibration framework which detects and corrects the overall systematic or bias error(s) introduced by both PMU and its instrumentation channel. Major contribution of the proposed method lies in that it does not require accurate prior knowledge of the system mathematical model/parameters. Furthermore, one byproduct of the proposed method is more accurate impedance parameters of the transmission line for EMS database and protective relay settings. By relaxing those strong assumptions employed in existing model based approaches, the proposed method advances the practicability of online PMU calibration.

The remainder of this paper is organized as follows. Section II describes the problem and related mathematical models. Section III presents the proposed framework. Case studies are presented in section IV while conclusion and future work are discussed in section V.

## II. PROBLEM DESCRIPTION AND FORMULATION

### A. Bias Error in PMU Measurements

Generally speaking, errors in synchrophasor measurements can originate from three possible sources as discussed in [19]: transducers, synchronization, and phasor estimation algorithm. Impacts of these three sources can be summarized into two categories: random error and bias (systematic) error.

Random error, as its name suggests, is random in either direction in its nature and difficult to predict. Random error can be circumvented from measurements via statistical means. Extensive studies have been conducted in reducing random error or its influences to PMU measurements, with satisfactory results observed: unbiased linear least squares (LS) is used in [21]; non-linear LS algorithms are used in [22]-[23]; total LS is introduced in [24]; other optimization procedures are discussed in [25] and [26].

Systematic or bias error is reproducible inaccuracy that is consistent in the same direction. Bias error is much harder to estimate and remove. Authors of [2] have examined the maximum bias errors introduced by different portions of the measurement chain. Table I summarizes the maximum bias error for a typical 230-kV system. For example, with a 400-ft instrumentation cable, the maximum bias errors in the magnitude and phase angle of the voltage phasor are 0.709% and 1.471 degree, respectively. These bias errors are no longer negligible and a systematic approach needs to be developed to identify and remove them, which is the scope of this paper.

TABLE I
MAXIMUM BIAS ERRORS INTRODUCED BY DIFFERENT PORTIONS OF THE PMU MEASUREMENT CHAIN IN A 230-KV SYSTEM [2]

| Source of Error | Voltage Phasor | | | | Current Phasor | | | |
|---|---|---|---|---|---|---|---|---|
| | Magnitude (%) | | Phasor Angle (degree) | | Magnitude (%) | | Phasor Angle (degree) | |
| PT/CT | 0.6 | | 1.04 | | 1.2 | | 0.52 | |
| | *100 ft* | *400 ft* | *100 ft* | *400 ft* | *100 ft* | *400 ft* | *100 ft* | *400 ft* |
| Cabling | 0.009 | 0.009 | 0.115 | 0.411 | 0.066 | 0.066 | 0.03 | 0.09 |
| PMU | 0.1 | | 0.02 | | 0.1 | | 0.02 | |
| Total | 0.709 | 0.709 | 1.175 | 1.471 | 1.366 | 1.366 | 0.57 | 0.63 |

### B. Notations and Models

Fig. 1 shows a measured voltage phasor $\bar{V}$, its corresponding true phasor value $\bar{V}_{true}$, and the associated bias errors in magnitude $\partial V$ and phase angle $\partial \theta$. The following relationship is derived:

$$\bar{V} = V \cdot e^{j\theta_V} \quad (1)$$

$$\bar{V}_{true} = (V + \partial V) \cdot e^{j(\theta_V + \partial \theta)} \quad (2)$$

where $V$ and $\theta_V$ are the magnitude and phase angle of phasor $\bar{V}$, respectively.

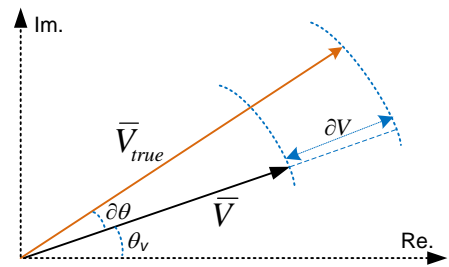

Fig. 1. Bias error in PMU measurement

A PI model as shown in Fig. 2 is considered in this work as the model for a general transmission line. It could either be a nominal PI model if the line is short or an equivalent PI if the line is longer [28]. In Fig. 2, $\bar{V}_s$ and $\bar{I}_s$ represent the positive sequence voltage and current phasors measured at sending end of the line while $\bar{V}_r$ and $\bar{I}_r$ are the corresponding phasors collected from receiving end. Variables $Z$ and $Y$ represent series impedance and shunt admittance of the PI model.

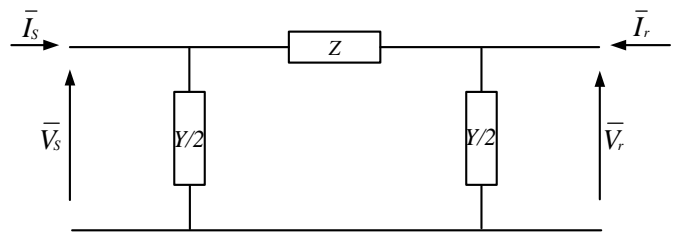

Fig. 2. Transmission line nominal/equivalent PI model

The following equations are derived from nodal analysis:

$$\bar{I}_s - \bar{V}_s \cdot \frac{Y}{2} + \bar{I}_r - \bar{V}_r \cdot \frac{Y}{2} = 0 \quad (3)$$

$$\bar{V}_s - Z \cdot \left(\bar{I}_s - \bar{V}_s \cdot \frac{Y}{2}\right) - \bar{V}_r = 0 \quad (4)$$

and

$$Z = R + jX \quad (5)$$

$$Y = G + jB_c \quad (6)$$

where $G$ and $B_c$ are line shunt conductance and susceptance. Combing equations (3)-(4) to solve for $Y$ and $Z$ yields:

$$Z = \frac{\bar{V}_s^2 - \bar{V}_r^2}{\bar{I}_s \cdot \bar{V}_r - \bar{I}_r \cdot \bar{V}_s} \quad (7)$$

$$Y = 2 \cdot \frac{\bar{I}_s + \bar{I}_r}{\bar{V}_s + \bar{V}_r} \qquad (8)$$

Substituting each phasor in (7)-(8) with its magnitude and phase angle according to (1) and setting phase angle of $I_r$, $\theta_{I_r}$, as the reference, the following equations can be derived:

$$R = real\left(\frac{V_s^2 e^{i2\theta'_{V_s}} - V_r^2 e^{i2\theta'_{V_r}}}{I_s \cdot V_r e^{i(\theta'_{I_s}+\theta'_{V_r})} - I_r \cdot V_s e^{i\theta'_{V_s}}}\right) \qquad (9)$$

$$X = imag\left(\frac{V_s^2 e^{i2\theta'_{V_s}} - V_r^2 e^{i2\theta'_{V_r}}}{I_s \cdot V_r e^{i(\theta'_{I_s}+\theta'_{V_r})} - I_r \cdot V_s e^{i\theta'_{V_s}}}\right) \qquad (10)$$

$$B_c = imag\left(2 \cdot \frac{I_s e^{i\theta'_{I_s}} + I_r}{V_s e^{i\theta'_{V_s}} + V_r e^{i\theta'_{V_r}}}\right) \qquad (11)$$

where $\theta'_{V_s} = \theta_{V_s} - \theta_{I_r}$, $\theta'_{V_r} = \theta_{V_r} - \theta_{I_r}$, and $\theta'_{I_s} = \theta_{I_s} - \theta_{I_r}$.

Shunt conductance of a transmission line, $G$, is usually very small and therefore neglected from the PI model.

### III. PROPOSED SOLUTION

#### A. Least Square Estimator (LSE)

To investigate the sensitivity of line parameters to bias errors in PMU measurements, partial derivatives need to be taken for (9)-(11), all of which are complex equations. For the derivatives to be valid, they must obey the Cauchy-Riemann equations [27]. The compliance checking/procedure is not discussed here due to space consideration, but the validation has been completed. The following equations have been derived:

$$\partial R = A_R \cdot \partial V_s + B_R \cdot \partial V_r + C_R \cdot \partial I_s + D_R \cdot \partial I_r + E_R \cdot \partial \theta'_{V_s} + F_R \cdot \partial \theta'_{V_r} + G_R \cdot \partial \theta'_{I_s} \qquad (12)$$

$$\partial X = A_X \cdot \partial V_s + B_X \cdot \partial V_r + C_X \cdot \partial I_s + D_X \cdot \partial I_r + E_X \cdot \partial \theta'_{V_s} + F_X \cdot \partial \theta'_{V_r} + G_X \cdot \partial \theta'_{I_s} \qquad (13)$$

$$\partial B_c = A_B \cdot \partial V_s + B_B \cdot \partial V_r + C_B \cdot \partial I_s + D_B \cdot \partial I_r + E_B \cdot \partial \theta'_{V_s} + F_X \cdot \partial \theta'_{V_r} + G_X \cdot \partial \theta'_{I_s} \qquad (14)$$

where coefficients $A_x \sim G_x$ are all partial derivatives. Taking $R$ as an example, these coefficients are: $A_R = \frac{\partial R}{\partial V_s}$, $B_R = \frac{\partial R}{\partial V_r}$, $C_R = \frac{\partial R}{\partial I_s}$, $D_R = \frac{\partial R}{\partial I_r}$, $E_R = \frac{\partial R}{\partial \theta'_{V_s}}$, $F_R = \frac{\partial R}{\partial \theta'_{V_r}}$, and $G_R = \frac{\partial R}{\partial \theta'_{I_s}}$. For space consideration, detailed information of these partial derivatives is discussed in Appendix I. Put equations (12)-(14) into matrix form to obtain:

$$\begin{bmatrix} \partial R \\ \partial X \\ \partial B_c \end{bmatrix} = \begin{bmatrix} A_R & B_R & C_R & D_R & E_R & F_R & G_R \\ A_X & B_X & C_X & D_X & E_X & F_X & G_X \\ A_B & B_B & C_B & D_B & E_B & F_B & G_B \end{bmatrix} \cdot \begin{bmatrix} \partial V_s \\ \partial V_r \\ \partial I_s \\ \partial I_r \\ \partial \theta'_{V_s} \\ \partial \theta'_{V_r} \\ \partial \theta'_{I_s} \end{bmatrix} \qquad (15)$$

It should be noted that coefficients $A_x \sim G_x$ vary with the loading (current), as can be seen from the expression of, for example, $C_R + iC_X = \frac{\partial Z}{\partial I_s} = \frac{V_r e^{i(\theta'_{V_r}+\theta'_{I_s})} \cdot (V_r^2 e^{i2\theta'_{V_r}} - V_s^2 e^{i2\theta'_{V_s}})}{\left(I_r V_s e^{i\theta'_{V_s}} - I_s V_r e^{i(\theta'_{V_r}+\theta'_{I_s})}\right)^2}$.

Assuming $N$ sets of PMU measurements under different load conditions are collected, the following matrix can be written:

$$H = \begin{bmatrix} A_{R1} & B_{R1} & C_{R1} & D_{R1} & E_{R1} & F_{R1} & G_{R1} \\ A_{X1} & B_{X1} & C_{X1} & D_{X1} & E_{X1} & F_{X1} & G_{X1} \\ A_{B1} & B_{B1} & C_{B1} & D_{B1} & E_{B1} & F_{B1} & G_{B1} \\ A_{R2} & B_{R2} & C_{R2} & D_{R2} & E_{R2} & F_{R2} & G_{R2} \\ A_{X2} & B_{X2} & C_{X2} & D_{X2} & E_{X2} & F_{X2} & G_{X2} \\ A_{B2} & B_{B2} & C_{B2} & D_{B2} & E_{B2} & F_{B2} & G_{B2} \\ \vdots & \vdots & \vdots & \vdots & \vdots & \vdots & \vdots \\ A_{RN} & B_{RN} & C_{RN} & D_{RN} & E_{RN} & F_{RN} & G_{RN} \\ A_{XN} & B_{XN} & C_{XN} & D_{XN} & E_{XN} & F_{XN} & G_{XN} \\ A_{BN} & B_{BN} & C_{BN} & D_{BN} & E_{BN} & F_{BN} & G_{BN} \end{bmatrix} \qquad (16)$$

$$F = \begin{bmatrix} \partial V_s \\ \partial V_r \\ \partial I_s \\ \partial I_r \\ \partial \theta'_{V_s} \\ \partial \theta'_{V_r} \\ \partial \theta'_{I_s} \end{bmatrix}, \quad E = \begin{bmatrix} \partial R_1 \\ \partial X_1 \\ \partial B_{c1} \\ \partial R_2 \\ \partial X_2 \\ \partial B_{c2} \\ \vdots \\ \partial R_n \\ \partial X_n \\ \partial B_{cn} \end{bmatrix} \qquad (17)-(18)$$

If an accurate set of line impedance parameters is known *a priori*, the bias error in the PMU measurements can be easily estimated using the standard least square estimator, as:

$$F = (H^T H)^{-1} H^T E \qquad (19)$$

Seven unknowns appear in $F$ and therefore the rank of $H$ matrix has to be no less than seven which requires $3 \times N \geq 7$ or $N \geq 3$, $(N \in N^*)$. Vector $E$ is comprised of the difference between the true line impedance values and the calculated ones using (9)-(11). With the assumption that accurate line impedances are known, here are the steps for evaluating the bias errors in PMU measurements:

- Step 1: calculate line impedance parameters $R$, $X$, and $B$ according to (9)-(11);
- Step 2: evaluate vector $E$ by comparing the calculated line impedances (from step1) to their corresponding references obtained from the EMS database: $R_{EMS}$, $X_{EMS}$, and $B_{EMS}$;
- Step 3: evaluate matrix $H$ with the partial derivatives calculated from PMU measurements: $\bar{V}_s$, $\bar{V}_r$, $\bar{I}_s$, and $\bar{I}_r$;
- Step 4: solve for vector $F$ based on (19).

The aforementioned least square estimator is able to identify bias errors in PMU measurements assuming the line's actual impedances, as the references, are known *a priori*. In practice, these parameters are read off from EMS database and were originally calculated based on tower geometries, conductor dimensions, estimates of line length, and conductor sags, etc. They only approximate the effects of conductor sags and ignore the dependence of impedance parameters on temperature and loading conditions [17]-[18]. Therefore, the challenge is that only approximates of line impedances are known and without knowing their true values the calculated bias errors might be far from being accurate. In the following section, this paper shows how this challenge can be addressed by using data mining technology like Density-based Spatial Clustering.

### B. Sensitivity Analysis

This subsection conducts sensitivity analysis and investigates influence of errors in referenced line impedances on bias error estimation in PMU measurements. A simulated transmission line with specifications shown in Appendix II is used for this study. Errors are added to all three line impedance references, one at a time, and the least square estimator described in section II. C is employed to evaluate $F$. As an example, results of the sensitivity analysis for line reactance $X$ are shown in Fig. 3-Fig. 4.

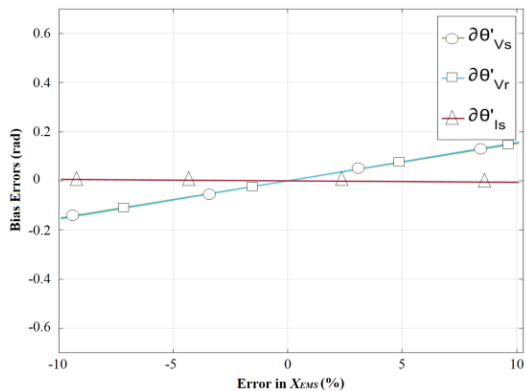

Fig. 3. Sensitivity analysis-estimated bias error in phase angle vs. line reactance

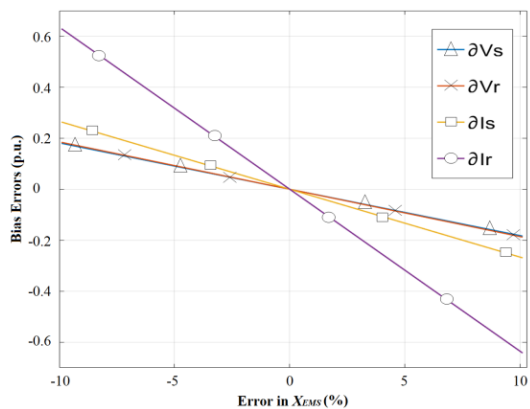

Fig. 4. Sensitivity analysis-estimated bias error in magnitude vs. line reactance

From Fig. 3-Fig. 4, it can be observed that the influence of error in $X_{EMS}$ on bias error estimation is linear. A 10% error in $X_{EMS}$ will result in: 0.16 rad error in $\partial\theta'_{V_s}$ and $\partial\theta'_{V_r}$; 0.01 rad error in $\partial\theta'_{I_s}$; 0.66 pu error in $\partial I_r$; 0.28 pu error in $\partial I_s$; 0.20 pu error in both $\partial V_s$ and $\partial V_r$. Bias error estimation in $\theta'_{I_s}$ is not really affected by error in $X_{EMS}$ while its corresponding magnitude is affected the most among all bias errors. Results of the sensitivity analysis are summarized in Table II.

A few observations can be made based on Table II, 1) PMU measurement bias error(s) estimation is generally sensitive to error(s) in line impedance references; 2) impact of reference errors to bias error estimation is linear; 3) if line impedance references are exactly known *a priori*, all bias error can be accurately estimated.

TABLE II
RESULTS FOR SENSITIVITY ANALYSIS

| | Error (%) | $\partial V_s$ | $\partial V_r$ | $\partial I_s$ | $\partial I_r$ | $\partial\theta'_{V_s}$ | $\partial\theta'_{V_r}$ | $\partial\theta'_{I_s}$ |
|---|---|---|---|---|---|---|---|---|
| | | p.u. | | | | rad | | |
| $\partial R$ | -10 | 0.051 | 0.051 | 0.072 | 0.152 | 0.005 | 0.005 | 0.001 |
| | -5 | 0.025 | 0.025 | 0.036 | 0.075 | 0.003 | 0.003 | 0.000 |
| | 0 | 0.000 | 0.000 | 0.000 | 0.000 | 0.000 | 0.000 | 0.000 |
| $\partial X$ | -10 | 0.200 | 0.200 | 0.283 | 0.658 | 0.163 | 0.162 | 0.010 |
| | -5 | 0.101 | 0.101 | 0.142 | 0.329 | 0.082 | 0.081 | 0.005 |
| | 0 | 0.000 | 0.000 | 0.000 | 0.000 | 0.000 | 0.000 | 0.000 |
| $\partial B_c$ | -10 | 0.061 | 0.061 | 0.162 | 0.360 | 0.090 | 0.080 | 0.010 |
| | -5 | 0.030 | 0.030 | 0.081 | 0.181 | 0.045 | 0.040 | 0.005 |
| | 0 | 0.000 | 0.000 | 0.000 | 0.000 | 0.000 | 0.000 | 0.000 |

However, in practice, line impedance references cannot be known exactly, as discussed in II. D. To address this challenge, a data mining approach based on DBSCAN is proposed to address the uncertainties in TL impedance references.

### C. DBSCAN Basics

Density-based spatial clustering of applications with noise (DBSCAN) is an unsupervised data mining technique which is able to classify data points of any dimension into core points, reachable points and outliers [20]. A core point $p$ contains at least *minPts* points (including $p$) within the designated searching distance $\varepsilon$. A reachable point $q$ exists if there exists a path $p1, p2,..., q$, so that all points on the path, except $q$, are core points. Points that are not reachable from any other point are outliers. Core points and reachable points can form a cluster while outliers are excluded from such cluster. Fig. 5 shows an example for DBSCAN with *minPts*=4. Note that setting *minPts* to 3 will generate the same clustering result.

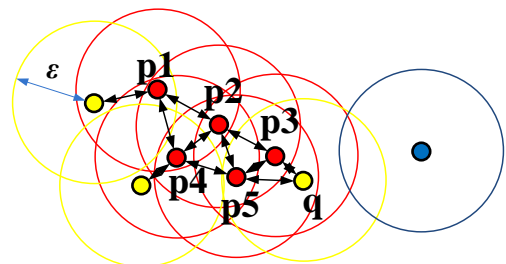

Fig. 5. Schematic diagram of DBSCAN with *minPts*=4

As shown in Fig. 5, core points are in red, each of which has at least 4 points with distance less than $\varepsilon$. The yellow ones (reachable points) are reachable from the red ones but do not have the required minimum number of points nearby within the distance of $\varepsilon$. The blue one is not reachable from any other point and therefore is an outlier. The red and yellow points form a cluster with the blue one excluded.

### D. PMU Calibration Using DBSCAN

Although EMS references can be significantly wrong, our experience shows that the error bands are generally well within 20%. Therefore, we may define $\alpha$ as the error band multiplier for impedance references obtained from the EMS database

($R_{EMS}$, $X_{EMS}$, and $B_{EMS}$). The following constraints can be considered:

$$\begin{cases} (1-\alpha)R_{EMS} \leq R \leq (1+\alpha)R_{EMS} \\ (1-\alpha)X_{EMS} \leq X \leq (1+\alpha)X_{EMS} \\ (1-\alpha)B_{EMS} \leq B_c \leq (1+\alpha)B_{EMS} \end{cases} \quad (20)$$

The corresponding feasibility region can be visualized as the cube shown in Fig. 6.

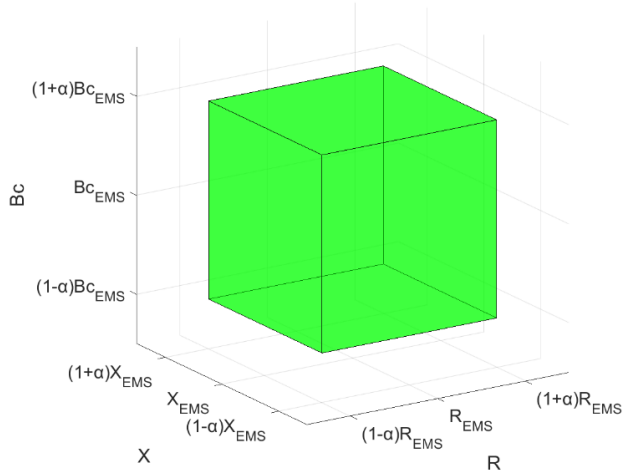

Fig. 6. Feasible region for transmission line (TL) impedances

The basic idea of the proposed approach is to 1) scan every point within feasible region (a total of $M$ points); 2) evaluate the corresponding bias errors in the PMU measurements; 3) form sets of points with each set containing seven 4-dimentional data points, and each data point has the form of $(\partial R, \partial X, \partial B_c, x)$, where $x$ is one of the bias errors in PMU measurements (a total of $M$ sets). 4) apply DBSCAN to cluster all the $M$ data sets to find out the one with least number of outliers (maximum number of core and reachable points) and minimum searching distance. Once this cluster is identified, the actual bias errors in all PMU channels and errors in line impedance references can be determined accordingly.

To minimize the computation, equation (19) is extended to (21). As compared to vector $E$ and $F$ in (19), matrix $E'$ (3$N$-by-$M$) and $F'$ (7-by-$M$) are the extended version which relates multiple bias error sets to multiple sets of the error in referenced impedances. Fig. 7 shows a simple illustrative example of the DBSCAN algorithm for bias error identification.

$$\begin{bmatrix} \partial V_s^1 & \partial V_s^2 & & \partial V_s^M \\ \partial V_r^1 & \partial V_r^2 & & \partial V_r^M \\ \partial I_s^1 & \partial I_s^2 & \cdots & \partial I_s^M \\ \partial I_r^1 & \partial I_r^2 & & \partial I_r^M \\ \partial \theta V_s'^1 & \partial \theta V_s'^2 & \cdots & \partial \theta V_s'^M \\ \partial \theta V_r'^1 & \partial \theta V_r'^2 & & \partial \theta V_r'^M \\ \partial \theta I_s'^1 & \partial \theta I_s'^2 & & \partial \theta I_s'^M \end{bmatrix} = (H^T H)^{-1} H^T \begin{bmatrix} \partial R_1^1 & \partial R_1^2 & & \partial R_1^M \\ \partial X_1^1 & \partial X_1^2 & & \partial X_1^M \\ \partial B_{c1}^1 & \partial B_{c1}^2 & & \partial B_{c1}^M \\ \partial R_2^1 & \partial R_2^2 & & \partial R_2^M \\ \partial X_2^1 & \partial X_2^2 & \cdots & \partial X_2^M \\ \partial B_{c2}^1 & \partial B_{c2}^2 & & \partial B_{c2}^M \\ \vdots & \vdots & & \vdots \\ \partial R_N^1 & \partial R_N^2 & & \partial R_N^M \\ \partial X_N^1 & \partial X_N^2 & & \partial X_N^M \\ \partial B_{cN}^1 & \partial B_{cN}^2 & & \partial B_{cN}^M \end{bmatrix} \quad (21)$$

Or

$$F' = (H^T H)^{-1} H^T E'$$

Fig. 7 shows an exemplary DBSCAN result for one data set (one column in $F'$). Point zero is pre-defined as one of the core point and the cluster search always starts from zero. The minimum number requirement $minPts$ is set to be 3. In this particular example, the clustering results indicates that a total of 6 points are either core or reachable points while two points are identified to be outliers (the ones with $\partial I_s$ and $\partial \theta'_{I_s}$).

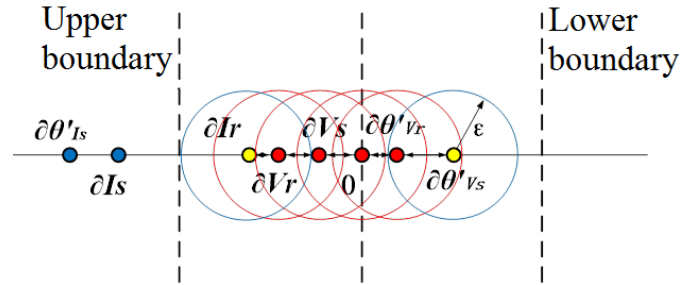

Fig. 7. An example of bias error identification using DBSCAN

The pseudo-code for the proposed DBSCAN based PMU calibration is presented below.

---
**algorithm** DBSCAN_based_PMU_calibration is
   **input:** PMU phasor measurements,
        Nodes ($R$, $X$, $B_c$) in the cube.
   **output:** PMU bias errors
    matrix($H$) ←PMU measurements
    matrix($E'$) ←Matrix(Nodes($R$, $X$, $B_c$) - EMS($R$, $X$, $B_c$))
    //(LSE: Least Square Estimator)
    matrix($F'$) ←LSE($H$, $E'$)
    vector(CCP)← //CCP: count cluster point
    $minPts$← 3
  **for** each column in matrix($F'$) **do**
  // $i$← row index for matrix ($F'$), (1≤$i$≤ 7)
  // $j$← column index for matrix ($F'$)
    residual($i$, $j$)← |$F'(i,j)$| (1≤$i$≤ 7)
    **if** residual($i$,$j$) is within the boundary $\alpha$
       $C(i,j) \leftarrow F'(i,j)$
       CCP($j$)++
  **Endforeach**
  **for** each column in matrix($C$) **do**
  // $i$← row index for matrix ($F'$), (1≤$i$≤ 7)
  // $j$← column index for matrix ($C$)
    distance($i$, $i$ ±1 ) ← |$C(i,j) - C(i \pm 1, j)$| . //find the distance from each cluster point to its neighbor cluster point
    eps($j$)= (maximum(|distance|), $minPts$) // eps($j$) is the cluster searching distance for $j^{th}$ column
  **Endforeach**
  $k$← column in $F'$ with maximum CPP and minimum eps
  **output**
---

A flowchart of the proposed data mining based PMU data calibration approach is shown in Fig. 8.

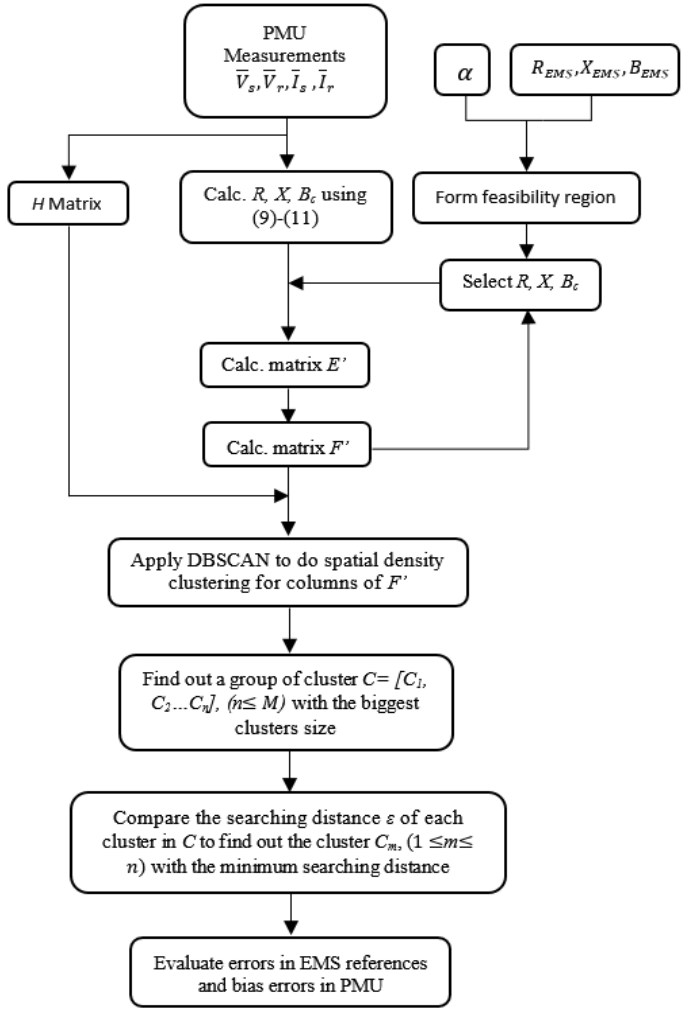

Fig. 8. Flow chart of proposed PMU data calibration approach

As shown in the sensitivity analysis of section III, impacts of EMS reference errors to the calculation of bias error in each PMU measurement are given. In the ideal case when there is no error in the EMS reference and no bias error in the PMU measurements, by using DBSCAN, all the seven curves in the 4-dimensional space ($\partial R, \partial X, \partial B_c, x$), where $x$ is one of the bias errors in PMU measurements, intersect at a single point (0, 0, 0, 0). When there are errors either in the EMS references or in the PMU measurements, according to the proposed approach, coordinates of this intersection corresponds to the error(s) in the EMS reference, while the outlier(s) identified correspondingly are the bias errors in PMU measurements.

The aforementioned discussion is for the situation without noise. When noise exists in PMU measurements, the seven curves typically will not intercept exactly at a single point but instead will stay very close to each other around one particular zone/region, which may be referred to as the "zero region". By looking at the searching distance and number of outliers, the "zero region" can be identified and therefore the errors in EMS references and PMU measurements can be evaluated accordingly.

## IV. EXPERIMENTAL VALIDATION

Four case studies are presented in this section to demonstrate the procedure and effectiveness of the proposed PMU data calibration framework. A testing system with parameters shown in Table VIII has been set up in Matlab/Simulink for these experiments.

### A. Case I: *The ideal case-no error in impedance references*

Objective of the first case study is to validate performance of the proposed method when no error exists in the TL impedance references. Basically, different sets of combinations of bias errors have been added to the PMU measurements and the proposed approach is used to identify them. The results for six representative cases are summarized in Table III, in which both the true bias errors and the calculated ones are presented and compared. The agreement between true bias errors and calculated ones validates the proposed approach under the ideal condition with no error in the referenced impedances.

TABLE III
BIAS ERROR IDENTIFICATION UNDER THE IDEAL CONDITION

|  | True | Calculated | True | Calculated | True | Calculated |
|---|---|---|---|---|---|---|
|  | p.u. *or* rad (×10⁻³) | | | | | |
| $\partial V_s$ | 0 | 0.0334 | 10 | 9.9634 | 10 | 10.0395 |
| $\partial V_r$ | 0 | 0.0335 | 10 | 9.9613 | 10 | 10.0402 |
| $\partial I_s$ | 10 | 10.0591 | 0 | -0.2240 | 10 | 10.0177 |
| $\partial I_r$ | 0 | 0.1480 | 0 | -0.5332 | 0 | -0.2318 |
| $\partial \theta'_{Vs}$ | 0 | 0.0067 | 1.75 | 1.7836 | 0 | 0.0185 |
| $\partial \theta'_{Vr}$ | 1.75 | 1.7601 | 0 | 0.0603 | 1.7 | 1.7536 |
| $\partial \theta'_{Is}$ | 0 | -0.0001 | 0 | -0.0070 | 0 | -0.0070 |
| $\partial V_s$ | 10 | 10.0161 | 10 | 10.0891 | 10 | 10.0033 |
| $\partial V_r$ | 10 | 10.0154 | 10 | 10.0894 | 10 | 10.0034 |
| $\partial I_s$ | 10 | 9.9647 | 10 | 10.0591 | 10 | 9.9328 |
| $\partial I_r$ | 0 | -0.3783 | 10 | 9.9659 | 10 | 9.6749 |
| $\partial \theta'_{Vs}$ | 0 | 0.0579 | 0 | 0.0138 | 1.7 | 1.7731 |
| $\partial \theta'_{Vr}$ | 1.75 | 1.7937 | 1.75 | 1.7522 | 1.7 | 1.7757 |
| $\partial \theta'_{Is}$ | 1.75 | 1.7663 | 1.75 | 1.7698 | 1.7 | 1.7676 |

### B. Case II: *One referenced impedance has error*

The second case study considers error in one of the referenced impedances. Towards this goal, errors are added to each of the referenced impedances, one at a time, and different combinations of bias errors are considered for PMU measurements. For space consideration, only the result for a representative case is presented below. And in this particular case, a -2% error is considered for the series resistance, $R_{EMS}$, and 0.01 p.u. bias error is added to magnitude of the sending-end current phasor. A 20% error band is considered for the referenced impedances with $\alpha$ being set to 20%. The proposed approach scans all 4-dimensional data points collected from matrix $F'$, and for visualization purpose, only the relationship between bias errors and errors in $R$ is plotted as shown in Fig. 9. The dashed line marks the outcome of DBSCAN and the X-axis gives the corresponding error in $R_{EMS}$. To help illustrate the DBSCAN process, Fig. 10 visualizes the corresponding DBSCAN clustering results, in terms of size of the cluster and searching distance. Final results are summarized in Table IV.

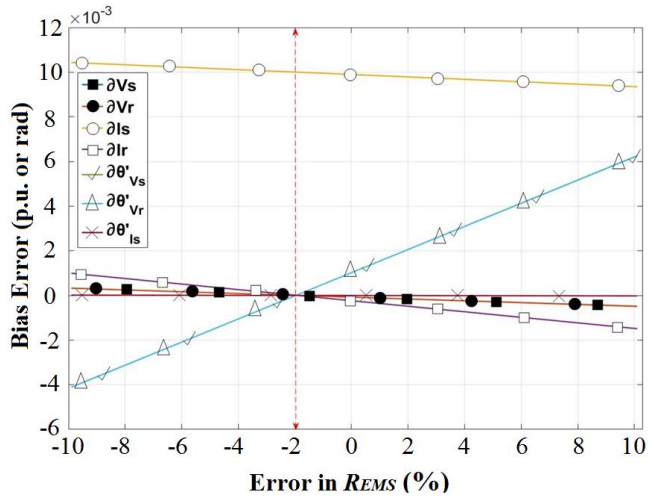

Fig. 9. Relationships between identified bias errors and error in $R_{EMS}$

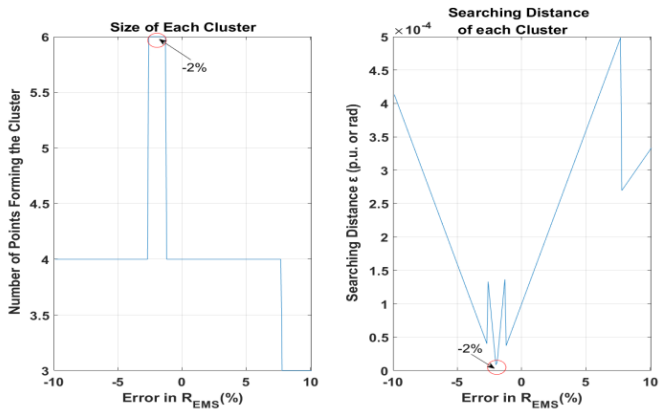

Fig. 10. Results of DBSCAN for case II

TABLE IV
TEST RESULTS FOR CASE II

|  | True | Calculated | Calculated Error in | | |
|---|---|---|---|---|---|
|  | p.u. *or* rad (×10⁻³) | | R (%) | X (%) | $B_C$ (%) |
| $\partial V_s$ | 0 | 0.0230 | | | |
| $\partial V_r$ | 0 | 0.0345 | | | |
| $\partial I_s$ | 10 | 10.0328 | | | |
| $\partial I_r$ | 0 | -0.0175 | -2.02 | 0.01 | -0.01 |
| $\partial \theta'_{Vs}$ | 0 | 0.0224 | | | |
| $\partial \theta'_{Vr}$ | 0 | 0.0130 | | | |
| $\partial \theta'_{Is}$ | 0 | -0.0149 | | | |

According to Table IV, the proposed approach successfully identifies not only bias error in PMU measurements but also error in the referenced TL series resistance.

### C. Case III: Two referenced impedances have errors

In the third case study, -4% error and -6% error are considered for $R_{EMS}$ and $X_{EMS}$, respectively; bias errors of 0.01 p.u. and 0.00175 rad are added to $V_s$ and $\theta_{Vr}$, respectively. A 20% error band is considered for the referenced impedances with $\alpha$ being set to 20%.

To help illustrate the DBSCAN process, Fig. 11 visualizes the clustering results, in terms of size of the cluster and searching distance. Both errors in the referenced impedances and the bias errors in PMU measurements are successfully identified. Final results are summarized in Table V.

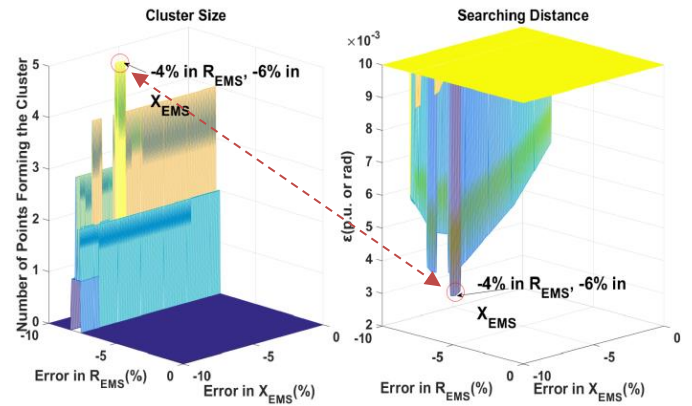

Fig. 11. DBSCAN clustering results for case III

TABLE V
TEST RESULTS FOR CASE III

|  | True | Calculated | Calculated Error in | | |
|---|---|---|---|---|---|
|  | p.u. *or* rad (×10⁻³) | | R (%) | X (%) | $B_C$ (%) |
| $\partial V_s$ | 10 | 10.0612 | | | |
| $\partial V_r$ | 0 | 0.0334 | | | |
| $\partial I_s$ | 0 | 0.0316 | | | |
| $\partial I_r$ | 0 | 0.1591 | −4.2 | −5.8 | 0.02 |
| $\partial \theta'_{Vs}$ | 0 | 0.0036 | | | |
| $\partial \theta'_{Vr}$ | 1.75 | 1.7617 | | | |
| $\partial \theta'_{Is}$ | 0 | 0.0047 | | | |

### D. Case IV: All referenced impedance values have errors

In the fourth case study, a set of -2%, -5%, 2% errors are considered for $R_{EMS}$, $X_{EMS}$, and $B_{EMS}$, respectively; bias errors of 0.01 p.u. and 0.00175 rad are added to $V_s$ and $\theta_{Vr}$, respectively. A 20% error band is considered for the referenced impedances with $\alpha$ being set to 20%. Experimental results shown in Table VI demonstrate again the effectiveness of the proposed method when all referenced impedance values have errors. One key value of the proposed approach lies in its capability of PMU calibration without knowing an accurate system model.

TABLE VI
TEST RESULTS FOR CASE IV

|  | True | Calculated | Calculated Error in | | |
|---|---|---|---|---|---|
|  | p.u. *or* rad (×10⁻³) | | R (%) | X (%) | $B_C$ (%) |
| $\partial V_s$ | 10 | 10.0143 | −2.1 | -5.2 | 2.2 |
| $\partial V_r$ | 0 | 0.0054 | | | |

| | | |
|---|---|---|
| $\partial I_s$ | 0 | -0.0375 |
| $\partial I_r$ | 0 | -0.0846 |
| $\partial\theta'_{Vs}$ | 1.75 | 1.7967 |
| $\partial\theta'_{Vr}$ | 0 | -0.0437 |
| $\partial\theta'_{Is}$ | 0 | -0.0091 |

*E. Case V: Experiment using real PMU data*

In the fifth case study, PMU data are collected for a 500-kV transmission line in Jiangsu Electricity Power Grid with the name of "*Huadong-Tianhui Line #5621*". PMU data reporting rate is 25 samples per second.

Using these real PMU data, the proposed approach is applied to identify both errors from EMS database and bias errors in the measured phasors. As discussed above, cluster size and searching distance, $\varepsilon$, are used as the criteria for DBSCAN. Part of the spatial clustering results are visualized in fig. 12, which shows the relationship between errors in the reference values of $R$, $X$, and $B_c$, and the cluster size. As shown in fig. 12, the top three points with the largest cluster size are *A1~A3*, of which *A2* has the smallest searching distance. The final results are summarized in Table VII. The results show,

1) TL impedance parameters, $R$, $X$ and $B_c$ in the EMS database have errors of -14%, 5.4% and 12.6% respectively;

2) no significant bias error in the voltage phasors collected from the real PMU is identified;

3) bias errors of 0.0171 pu and 0.0164 pu are identified in the magnitudes of sending-end and receiving-end current phasors, respectively;

4) no significant bias error is identified in the phase angles of the two current phasors.

TABLE VII
TEST RESULTS FOR CASE V

| | Calculated | Calculated Error in | | |
|---|---|---|---|---|
| | p.u. *or* rad (×10⁻³) | $R$ (%) | $X$ (%) | $B_C$ (%) |
| $\partial V_s$ | -0.0032 | | | |
| $\partial V_r$ | -0.0033 | | | |
| $\partial I_s$ | 0.0171 | -14.0 | 6.4 | 12.6 |
| $\partial I_r$ | 0.0164 | | | |
| $\partial\theta'_{Vs}$ | 0.0003 | | | |
| $\partial\theta'_{Vr}$ | 0.0027 | | | |
| $\partial\theta'_{Is}$ | -5.6238e-5 | | | |

Computation time of the proposed approach is dependent upon the number of data points to scan within the feasibility region. In one experiment, a total of 1 million points (worst case scenario) are processed using a Matlab program, and the solution process takes roughly 29 seconds (recoding the program using *C++* will greatly speed up the solution process). Fortunately, PMU data calibration does not need to be conducted very often, and once a week or longer will work for most cases. Structure of the proposed algorithm is suitable for parallel processing, which will further speed up the solution process.

V. CONCLUSION

This paper presents a novel approach for online calibration of PMU by using density-based spatial clustering. As compared to existing methods, the proposed one has two major merits: 1) it identifies the overall bias errors introduced by both PMU and its instrumentation channel; 2) it does not require accurate system model/parameters. Therefore, it is applicable across a wide spectrum of practical conditions. In addition, one by-product of the proposed approach is more accurate TL impedance estimates for improved system modeling, more accurate protective relay settings, and other related applications. Future work includes: 1) extending the proposed framework to system level to achieve simultaneous calibration of multiple PMUs; 2) decomposing the spatial clustering process so that state-of-the-art parallel computing techniques can be employed to speed up the computation.

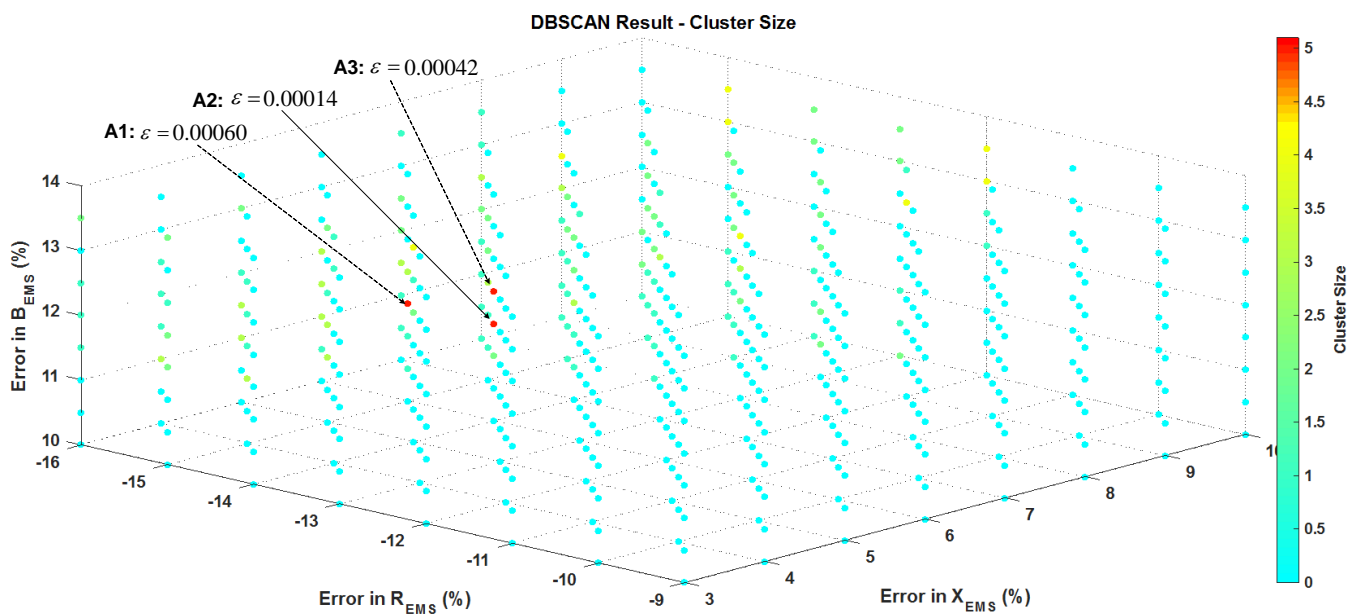

Fig. 12. Experiment result with real PMU data (errors in *R*, *X*, and *B_c* vs. cluster size)

## APPENDIX I

Partial derivatives of impedance to the PMU measurements are presented as follows:

$$\frac{\partial Z}{\partial V_s} = -\frac{2V_s e^{2i\theta'_{V_s}}}{I_r V_s e^{i\theta'_{V_s}} - I_s V_r e^{i(\theta'_{V_r}+\theta'_{I_s})}} \\ -\frac{I_r e^{i\theta'_{V_s}}(V_r^2 e^{2i\theta'_{V_r}} - V_s^2 e^{2i\theta'_{V_s}})}{(I_r V_s e^{i\theta'_{V_s}} - I_s V_r e^{i(\theta'_{V_r}+\theta'_{I_s})})^2} \quad (22)$$

$$\frac{\partial Z}{\partial V_r} = \frac{2V_r e^{2i\theta'_{V_r}}}{I_r V_s e^{i\theta'_{V_s}} - I_s V_r e^{i(\theta'_{V_r}+\theta'_{I_s})}} \\ +\frac{I_s e^{i(\theta'_{V_r}+\theta'_{I_s})}(V_r^2 e^{2i\theta'_{V_r}} - V_s^2 e^{2i\theta'_{V_s}})}{(I_r V_s e^{i\theta'_{V_s}} - I_s V_r e^{i(\theta'_{V_r}+\theta'_{I_s})})^2} \quad (23)$$

$$\frac{\partial Z}{\partial I_s} = \frac{V_r e^{i(\theta'_{V_r}+\theta'_{I_s})}(V_r^2 e^{2i\theta'_{V_r}} - V_s^2 e^{2i\theta'_{V_s}})}{(I_r V_s e^{i\theta'_{V_s}} - I_s V_r e^{i(\theta'_{V_r}+\theta'_{I_s})})^2} \quad (24)$$

$$\frac{\partial Z}{\partial I_r} = -\frac{V_s e^{i\theta'_{V_s}}(V_r^2 e^{2i\theta'_{V_r}} - V_s^2 e^{2i\theta'_{V_s}})}{(I_r V_s e^{i\theta'_{V_s}} - I_s V_r e^{i(\theta'_{V_r}+\theta'_{I_s})})^2} \quad (25)$$

$$\frac{\partial Z}{\partial \theta'_{V_s}} = -\frac{2iV_s^2 e^{2i\theta'_{V_s}}}{I_r V_s e^{i\theta'_{V_s}} - I_s V_r e^{i(\theta'_{V_r}+\theta'_{I_s})}} \\ -\frac{iI_r V_s e^{i\theta'_{V_s}}(V_r^2 e^{2i\theta'_{V_r}} - V_s^2 e^{2i\theta'_{V_s}})}{(I_r V_s e^{i\theta'_{V_s}} - I_s V_r e^{i(\theta'_{V_r}+\theta'_{I_s})})^2} \quad (26)$$

$$\frac{\partial Z}{\partial \theta'_{V_r}} = \frac{2iV_r^2 e^{2i\theta'_{V_r}}}{I_r V_s e^{i\theta'_{V_s}} - I_s V_r e^{i(\theta'_{V_r}+\theta'_{I_s})}} \\ -\frac{iI_s V_r e^{i(\theta'_{V_r}+\theta'_{I_s})}(V_r^2 e^{2i\theta'_{V_r}} - V_s^2 e^{2i\theta'_{V_s}})}{(I_r V_s e^{i\theta'_{V_s}} - I_s V_r e^{i(\theta'_{V_r}+\theta'_{I_s})})^2} \quad (27)$$

$$\frac{\partial Z}{\partial \theta'_{I_s}} = \frac{iI_s V_r e^{i(\theta'_{V_r}+\theta'_{I_s})}(V_r^2 e^{2i\theta'_{V_r}} - V_s^2 e^{2i\theta'_{V_s}})}{(I_r V_s e^{i\theta'_{V_s}} - I_s V_r e^{i(\theta'_{V_r}+\theta'_{I_s})})^2} \quad (28)$$

$A_R = real(\frac{\partial Z}{\partial V_s})$, $A_X = imag(\frac{\partial Z}{\partial V_s})$, $B_R = real(\frac{\partial R}{\partial V_r})$, $B_X = imag(\frac{\partial R}{\partial V_r})$, $C_R = real(\frac{\partial Z}{\partial I_s})$, $C_X = imag(\frac{\partial Z}{\partial I_s})$, $D_R = real(\frac{\partial Z}{\partial I_r})$, $D_X = imag(\frac{\partial Z}{\partial I_r})$, $E_R = real(\frac{\partial Z}{\partial \theta'_{V_s}})$, $E_X = imag(\frac{\partial Z}{\partial \theta'_{V_s}})$, $F_R = real(\frac{\partial Z}{\partial \theta'_{V_r}})$, $F_X = imag(\frac{\partial Z}{\partial \theta'_{V_r}})$, $G_R = real(\frac{\partial Z}{\partial \theta'_{I_s}})$, $G_X = imag(\frac{\partial Z}{\partial \theta'_{I_s}})$. By the same means, the partial derivative equations of $Y$ to each PMU components can be generated and $A_B, B_B \cdots G_B$ can be calculated accordingly. Due to the space limitation, they are not presented here.

## APPENDIX II

A transmission line with two PMUs installed at both terminals is simulated in this study using *Matlab/Simulink* with specifications shown in Table VIII.

TABLE VIII
SPECIFICATIONS OF THE SIMULATED TRANSMISSION LINE

| Variables | Description, Unit | Value |
|---|---|---|
| $R_{line}$ | line resistance, $\Omega/km$ | 0.013333 |
| $L_{line}$ | line inductances, $H/km$ | 7.4342e-4 |
| $C_{line}$ | line capacitances, $F/km$ | 1.0001e-8 |
| $D$ | length of line, $km$ | 150 |
| $f_{source}$ | source frequency, $Hz$ | 60 |
| *Voltage level* | $kV$ | 500 |

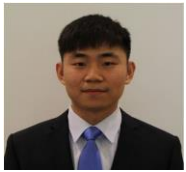
**Xinan Wang** (S'15) received the B.S. degree in electrical engineering from Northwestern Polytechnical University, Xi'an, China, in 2013, and the M.S. degree in electrical engineering from Arizona State University, Tempe, AZ, USA, in 2016. He currently works as a research assistant in the Advanced Power System Analytics Group at GEIRI North America. His research interests include WAMS and grid integration of renewables.

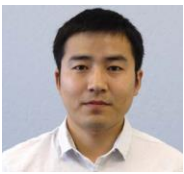
**Di Shi** (M'12, SM'17) received the Ph.D. degree in electrical engineering from Arizona State University, Tempe, AZ, USA, in 2012. He currently leads the Advanced Power System Analytics Group at GEIRI North America, Santa Clara, CA, USA. He has published over 50 journal and conference papers and hold 13 US patents/patent applications.

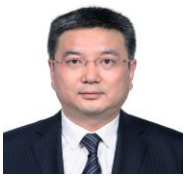
**Zhiwei Wang** received the B.S. and M.S. degrees in electrical engineering from Southeast University, Nanjing, China, in 1988 and 1991, respectively. He is President of GEIRI North America, Santa Clara, CA, USA. His research interests include power system operation and control, relay protection, power system planning, and WAMS.

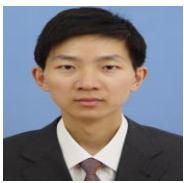
**Chunlei Xu** received the B.S. degree in electrical engineering from Shanghai Jiao Tong University, Shanghai, China, in 1999. He currently leads the Dispatching Automation department at Jiangsu Electrical Power Company in China. His research interests include power system operation and control and WAMS.

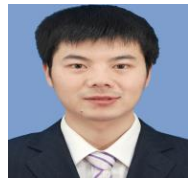
**Qibing Zhang** received the B.S. degree in electrical engineering from Zhejiang University, Zhejiang, China, in 2007, and the M.S. degree in electrical engineering from Shanghai Jiao Tong University, Shanghai, China, in 2010. His research interests include power system operation and control and relay protection.

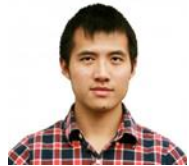
**Xiaohu Zhang** (S'12) received the B.S. degree in electrical engineering from Huazhong University of Science and Technology, Wuhan, China, in 2009, the M.S. degree in electrical engineering from Royal Institute of Technology, Stockholm, Sweden, in 2011, and the Ph.D. degree in electrical engineering at The University of Tennessee, Knoxville, in 2017.

Currently, he works as a power system engineer at GEIRI North America, Santa Clara, CA, USA. His research interests are power system operation, planning and stability analysis.

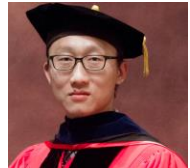
**Zhe Yu** received his B.E. degree from Department of Electrical Engineering, Tsinghua University, Beijing, China in 2009, M.S. degree from Department of Electrical and Computer Engineering, Carnegie Mellon University, Pittsburgh, PA, USA in 2010, and Ph.D. from the School of Electrical and Computer Engineering, Cornell University, Ithaca, NY, USA in 2016. He joined Globe Energy Internet Research Institute North America in 2017. His current research interests focus on the power system and smart grid, demand response, dynamic programming, and optimization.